\documentclass[aps,prc,reprint,superscriptaddress]{revtex4-2}
\usepackage{ulem}
\usepackage{amsmath}
\usepackage{appendix}
\usepackage{url}
\usepackage[colorlinks=true,linkcolor=blue,filecolor=blue, urlcolor=blue,citecolor=blue]{hyperref}
\usepackage{float}
\usepackage[T1]{fontenc}
\usepackage{graphicx}
\usepackage{mathrsfs}
\usepackage{dcolumn}
\usepackage{bm}
\usepackage{cases}
\usepackage{multirow,booktabs}
\usepackage{makecell}
\usepackage{amssymb}
\usepackage{eucal}
\usepackage{subfigure}
\usepackage{natbib}
\usepackage{amsthm,amsmath,amssymb}
\usepackage{mathrsfs}
\usepackage{subeqnarray}
\usepackage{color}
\usepackage{ulem}
\usepackage{soulutf8}
\usepackage{appendix}
\usepackage{listings}
\usepackage{braket}
\usepackage{multirow,booktabs}
\usepackage{chngcntr}

\begin{document}
	
\title{Nuclear contacts of unstable nuclei}

\author{Tongqi Liang}
\affiliation{School of Physics Science and Engineering, Tongji University, Shanghai 200092, China}

\author{Dong Bai}
\affiliation{College of Mechanics and Engineering Science, Hohai University, Nanjing 211100, China}
	
\author{Zhongzhou Ren}
\email{zren@tongji.edu.cn}
\affiliation{School of Physics Science and Engineering, Tongji University, Shanghai 200092, China}
\affiliation{Key Laboratory of Advanced Micro-Structure Materials, Ministry of Education, Shanghai 200092, China}

\begin{abstract}
Nuclear contact is a key quantity to describe the nucleon-nucleon short-range correlations (SRCs). While they have been determined by electron scattering experiments for selected stable nuclei, nuclear contacts are largely unknown for unstable nuclei.
In this work, we study nuclear contacts for a number of nuclei in the vicinity of the doubly magic $^{132}$Sn from the theoretical perspective, with special emphasis on unstable nuclei. 
We find that the proton-proton contact generally gets suppressed by the excess neutrons for the Sn isotopes,
 resembling the suppression of $\alpha$-cluster formation reported recently for the same isotopic chain [J.\ Tanaka {\it{et al.}}, Science {\bf 371}, 260 (2021)].
This indicates a hidden universal aspect of SRCs and $\alpha$ clustering, two different kinds of nuclear correlations. 
Meanwhile, a linear relation is found between the proton-proton contact and the proton number for the $N=82$ isotones.
Our results can be helpful for future experimental studies of SRCs in unstable nuclei at advanced facilities worldwide.

\end{abstract}

\maketitle
	
\section{ introduction}

One of the main challenges in describing the nuclear system is understanding the nucleon-nucleon ($NN$) short-range correlations (SRCs). The contact, first introduced by Tan in the ultra-cold atomic system, 
represents the probability of finding two unlike fermions being in close proximity, serving as a powerful tool to investigate correlated fermions via short-range forces~\cite{TAN20082952,TAN20082971,TAN20082987}. The contact theory has been successfully generalized to the nuclear system by applying the generalized contact formalism (GCF), and significant progress has been made in nuclear contacts~\cite{Weiss2015,Weiss2017,Schmidt2020,Cruz-Torres2020}. Different from the atomic system, there are several kinds of correlated pairs in nuclei, and one needs to account for the complex nature of the nuclear force. By applying GCF, pair abundances in nuclei have been addressed and have shown its adhibition in modeling the effects of SRCs on neutron stars~\cite{Hen2015,cai2016,Hong2023} and in studies of the European Muon Collaboration (EMC) effects~\cite{Weinstein2011,Hen2012,Hen2017,chen2017}. 

It is widely known that the correlated nucleons being very close have relatively high momentum, which provides an intuitive connection between the nuclear contact and the high-momentum tails of the nucleon momentum distribution~\cite{Day1990,Fomin2012,Ivanov2020,LYU2020,Liang2022,Wang2023,Liang2024}. With this relation, the nuclear contacts of distinct nucleon pairs for chosen nuclei are addressed using the \textit{ab initio} calculations and quasi-elastic electron scattering experiments~\cite{Weiss2018}. Nevertheless, \textit{ab initio} calculations are currently limited to light and medium-size nuclei, and quasi-elastic electron scattering experiments can only be performed for chosen stable nuclei. 

The study of short-lived unstable nuclei has garnered great interest due to their association with intriguing nuclear phenomena such as halos, proton bubbles, and variations in nuclear magic numbers~\cite{GARRIDO1999,Grasso,SUDA2017,Liu2017,Liang2018,Wang2020}. 
With the development of radioactive nuclear beam facilities, elastic electron scattering experiments off unstable nuclei have become available at RIKEN~\cite{Tsukada2017,Tsukada2023}. 
A deep comprehension of the exotic features is crucial not only for advancing nuclear structure theories but also for understanding nucleosynthesis processes that play a pivotal role in astrophysics~\cite{KAJINO2019}.
In Ref.~\cite{Weiss2019}, authors found that in the framework of GCF, nuclear SRCs are closely related to the nuclear charge density, i.e., the one-body proton density. Without an explicit consideration of SRCs, the nuclear charge density can be determined from elastic electron scattering and mean-field calculations, making it possible to extend the investigation of nuclear contacts to unstable nuclei.

In this work, we use the proton densities, given by the Skyrme Hartree-Fock-Bogolyubov (HFB) model, to extract the $pp$ contact for Sn isotopes and $N$=82 isotones. These two chains encompass many stable and unstable nuclei. These nuclei are also being actively researched at RIKEN~\cite{Ohnishi2023}. We identify correlations between the $pp$ contact and other nuclear properties, such as the root-mean-square (RMS) radius of the proton density distributions $r_p$ and the effective number of $\alpha$ clusters.

\section{ theoretical framework}
\subsection{Generalized contact formalism}\label{contact_theo}

The original contact theory given by Tan was formulated for ultracold atomic systems with significant scale separation. When two particles get very close, the many-body wave function $\Psi$ can be factorized into a product of an asymptotic pair wave function $\varphi(\textit{\textbf{r}}_{ij})$ and a function $A_{ij}$ describing the residual $A-2$ system~\cite{TAN20082952,TAN20082971,TAN20082987,Werner2012}
\begin{equation}
    \Psi \underset{r_{i j} \rightarrow 0}{\longrightarrow} \varphi(\textit{\textbf{r}}_{ij}) A_{ij}(\textit{\textbf{R}}_{ij},\{\textit{\textbf{r}}_k\}_{k\neq i,j}).
\end{equation}
where $\textit{\textbf{r}}_{ij}$ and $\textit{\textbf{R}}_{ij}$ are respectively the relative and center-of-mass coordinate of the pair.
For the system satisfying the scale separation, the short-range interaction can be replaced by a boundary condition and all partial waves are suppressed but $s$-wave. In such case, the asymptotic pair wave function is expressed as $\varphi(\textit{\textbf{r}}_{ij})=(1/r_{ij}-1/a_s)$ with $r_{ij}=|\textit{\textbf{r}}_{ij}|$ and $a_s$ is the scattering length. The contact $C$ is defined as 
\begin{equation}
    C=16\pi^2 \sum_{ij}\braket{A_{ij}|A_{ij}}.
\end{equation}

When the contact theory is generalized to the nuclear system, there are more kinds of pairs consisting of protons and neutrons with spin being either up or down. Moreover, the assumption of $s$-wave dominance does not hold for nuclei~\cite{Weiss2017,Cruz2018}. Consequently, in the framework of GCF, all possible $NN$ channels $\alpha$ must be included, and the many-body wave function takes the form
\begin{equation}
    \Psi \underset{r_{i j} \rightarrow 0}{\longrightarrow}\sum_{\alpha} \varphi_{\alpha}(\textit{\textbf{r}}_{ij}) A_{ij}^{\alpha}(\textit{\textbf{R}}_{ij},\{\textit{\textbf{r}}_k\}_{k\neq i,j}).
\end{equation}
Nuclear contacts are typically represented as matrices denoted by $C_{ij}^{\alpha \beta}$.  This work concentrates specifically on the diagonal elements $C_{ij}^{\alpha}$, which are directly proportional to the probability of identifying a nucleon-nucleon pair in close proximity in the $\alpha$ channel in the nucleus. The diagonal elements of nuclear contacts are defined as :
\begin{equation}
C^{\alpha}_{ij}=16\pi^2\sum_{ij}\braket{A^{\alpha}_{ij}|A^{\alpha}_{ij}}.
\end{equation}
The two most significant channels in nuclei are the spin-zero s-wave channel denoted by $\alpha=0$ and the spin-one deuteron channel denoted by $\alpha=1$~\cite{Weiss2019}, which is occupied only by proton-neutron pairs caused by the tensor force~\cite{Schiavilla2007}. The two-body functions $\varphi^{\alpha}(\textit{\textbf{r}}_{ij})$ are identical for all nuclei and are given by solving the zero-energy two-body problem. In this work, the AV18 potential is used to calculate $\varphi^{\alpha}(\textit{\textbf{r}}_{ij})$~\cite{Wiringa1995}.

At a small distance, the two-body density $\rho_{NN}^{\alpha}$, which defines the probability of finding a nucleon-nucleon pair with the separation distance $r=|\boldsymbol{r}|$, is related to SRCs and corresponds to the high-momentum tail in the nucleon momentum distribution. 
In GCF, $\rho_{NN}^{\alpha}(\textit{\textbf{r}})$ can be modeled by $\varphi^{\alpha}(\textit{\textbf{r}})$ and nuclear contacts
\begin{equation}\label{contact}
    \rho_{NN}^{\alpha}(\textit{\textbf{r}})=C_{NN}^{\alpha}|\varphi^{\alpha}(\textit{\textbf{r}})|^2.
\end{equation}
In the previous variational Monte Carlo (VMC) calculations, it was found that this relation holds for nuclei with $A\leq40$ in the range of $r<r_0\approx0.9$ fm~\cite{Weiss2018}.

At large separation distances, it is expected that two nucleons are not correlated, and the two-body density $\rho^{\alpha}_{NN}$ can be constructed by integrating the one-body point-nucleon density $\rho_N(\textit{\textbf{r}})$
\begin{equation}\label{onebody}
    \rho_{NN}(\boldsymbol{r}) \!\propto \!\rho_{NN}^{U C}(\boldsymbol{r})\! \equiv\! \int d \boldsymbol{R} \rho_{N}(\boldsymbol{R}+\boldsymbol{r} / 2) \rho_{N}(\boldsymbol{R}-\boldsymbol{r} / 2),
\end{equation}
Accounting for the fermionic nature of the $NN$ pair, the two-body density $\rho^{\alpha}_{NN}$ at large separation distances has the following form~\cite{Cruz2018}
\begin{equation}\label{twobody}
    \rho_{NN}(r) \underset{r \rightarrow \infty}{\longrightarrow} \rho_{NN}^{F}(r) \equiv \mathcal{N} \rho_{NN}^{U C}(r)\left[1-\frac{1}{2}\left(\frac{3 j_{1}\left(k_{F}^{} r\right)}{k_{F}^{} r}\right)^{2}\right],
\end{equation}
where $\mathcal{N}$ is the normalization, $j_1$ is the spherical Bessel function, and $k_F$ is the Fermi momentum of protons or neutrons. The above relation is deduced from the local-density approximation to the expansion of the density matrix~\cite{Negele1972}, which provides a good approximation for nuclear matter and heavy nuclei. 

Equations (\ref{onebody}) and (\ref{twobody}) provide an asymptotic expression for the two-body density at large separation distances, which can be directly obtained from the point-nucleon one-body density.
In Eq.~(\ref{contact}), we have shown that the two-body density at small separation distances can be well reproduced by the nuclear contact. It is verified that at the match point $r_0$, both the contact and $\rho_{NN}^{F}$ expressions describe the full two-body density well~\cite{Weiss2019}. This indicates the extraction of nuclear contact with the relation
\begin{equation}
    C_{NN}^{\alpha}=\frac{\rho_{NN}^{F}\left(r_{0}\right)}{\left|\varphi_{NN}^{\alpha}\left(r_{0}\right)\right|^{2}}.
\end{equation}
Based on the above formulas, the $NN$ contact can be obtained by using only the one-nucleon density distributions in coordinate space.

\subsection{Skyrme Hartree-Fock-Bogolyubov model}

In the cooordinate representation, the Skyrme Hartree-Fock-Bogolyubov (HFB) energy can be written as~\cite{Stoitsov2005}
\begin{equation}
    E[\rho, \tilde{\rho}]=\int d \boldsymbol{r} \mathcal{H}(\boldsymbol{r}),
\end{equation}
with the energy density $\mathcal{H}(\boldsymbol{r})$ being a sum of the particle energy density $H(\boldsymbol{r})$ and the pairing energy density $\tilde{H}(\boldsymbol{r})$. 
$H(\boldsymbol{r})$ and $\tilde{H}(\boldsymbol{r})$ depend on the local particle density $\rho(\boldsymbol{r})$ and local pairing density $\tilde{\rho}(\boldsymbol{r})$. The variation of the HFB energy with  respect to $\rho(\boldsymbol{r})$ and $\tilde{\rho}(\boldsymbol{r})$ leads to the Skyrme HFB equations
\begin{widetext}
\begin{equation}
\sum_{\sigma^{\prime}}\left(\begin{array}{cc}
h\left(\boldsymbol{r}, \sigma, \sigma^{\prime}\right) & \tilde{h}\left(\boldsymbol{r}, \sigma, \sigma^{\prime}\right) \\
\tilde{h}\left(\boldsymbol{r}, \sigma, \sigma^{\prime}\right) & -h\left(\boldsymbol{r}, \sigma, \sigma^{\prime}\right)
\end{array}\right)\binom{U_k\left(E_k, \boldsymbol{r} \sigma^{\prime}\right)}{V_k\left(E_k, \boldsymbol{r}\sigma^{\prime}\right)}=\left(\begin{array}{cc}
E_k+\lambda & 0 \\
0 & E_k-\lambda
\end{array}\right)\binom{U_k(E_k, \boldsymbol{r} \sigma)}{V_k(E_k, \boldsymbol{r} \sigma)},
\end{equation}
where $\sigma=\pm\frac{1}{2}$ represents the spin and $\lambda$ is the Lagrange multiplier to fix the correct particle number. 
The local particle field $h\left(\boldsymbol{r}, \sigma, \sigma^{\prime}\right)$ and the local pairing field $\tilde{h}\left(\boldsymbol{r}, \sigma, \sigma^{\prime}\right)$ are given by taking the variation of Skyrme HFB energy, and the explicit form can be found in Ref.~\cite{Stoitsov2005}.
By solving the Skyrme HFB equation, one can obtain the upper ($U_k$) and lower ($V_k$) components of the quasiparticle wave function corresponding to the positive quasiparticle energy $E_k$. 

For the nucleus with axial symmetry, the third component of the total angular momentum is conserved, which provides a good quantum number $\Omega_k$, and quasiparticle HFB wave functions can be written as 
\begin{equation}
\binom{U_k(E_k,\boldsymbol{r}, \sigma, \tau)}{V_k(E_k,\boldsymbol{r}, \sigma, \tau)}=\chi_{q_k}(\tau)\left[\binom{U_k^{+}(r, z)}{V_k^{+}(r, z)} \mathrm{e}^{\mathrm{i} \Lambda^{-} \varphi} \chi_{+1 / 2}(\sigma)+\binom{U_k^{-}(r, z)}{V_k^{-}(r, z)} \mathrm{e}^{\mathrm{i} \Lambda^{+} \varphi} \chi_{-1 / 2}(\sigma)\right],
\end{equation}
\end{widetext}
where $\Lambda=\Omega_k\pm1/2$. $r,z,$ and $\phi$ are the standard cylindrical coordinates with the relation $\boldsymbol{r}=(r\cos \phi, r\sin \phi, z)$. $\chi_{q_k}(\tau)$ and $\chi_{\pm1 / 2}(\sigma)$ are isospin and spin functions, respectively~\cite{Sarriguren2007}.
The axially deformed one-nucleon density distributions in coordinate space is then defined as 
\begin{equation}
    \rho(r, z)=\sum_k\left(\left|V_k^{+}(r, z)\right|^2+\left|V_k^{-}(r, z)\right|^2\right).
\end{equation}

\section{Numerical results}
Within the framework of GCF, we extract the nuclear contacts of the spin-zero s-wave channel for $pp$ pairs, denoted by $C^{0}_{pp}$, for the Sn isotopes and $N$=82 isotones with very long chains, which provides the unique opportunity to investigate both the stable and unstable nuclei systematically. For these nuclei, the \textit{ab initio} method is unable to provide two-body densities. In this work, the Skyrme HFB model is applied to provide the corresponding point-proton density and further the two-proton density $\rho^0_{pp}$. Furthermore, the nuclear contact $C^{0}_{pp}$ is calculated in the coordinate space. The universal two-body function $\varphi^{0}_{pp}(r)$ is calculated by solving the two-nucleon Schrodinger equation for zero energy with the AV18 potential. $\varphi^{0}_{pp}(r)$ is normalized such that the integral of $\tilde{\varphi}^{0}_{pp}(k)$, the Fourier transform of $\varphi^{0}_{pp}(r)$, above the Fermi momentum equals unity~\cite{Cruz-Torres2020}. The error of $C^{0}_{pp}$ is expected to be of the order of 10$\%$, originating from the use of the infinite nuclear-matter approximation and the selection of $r_0$~\cite{Weiss2019}.

\begin{figure}
    \centering
    \includegraphics[width=1\columnwidth]{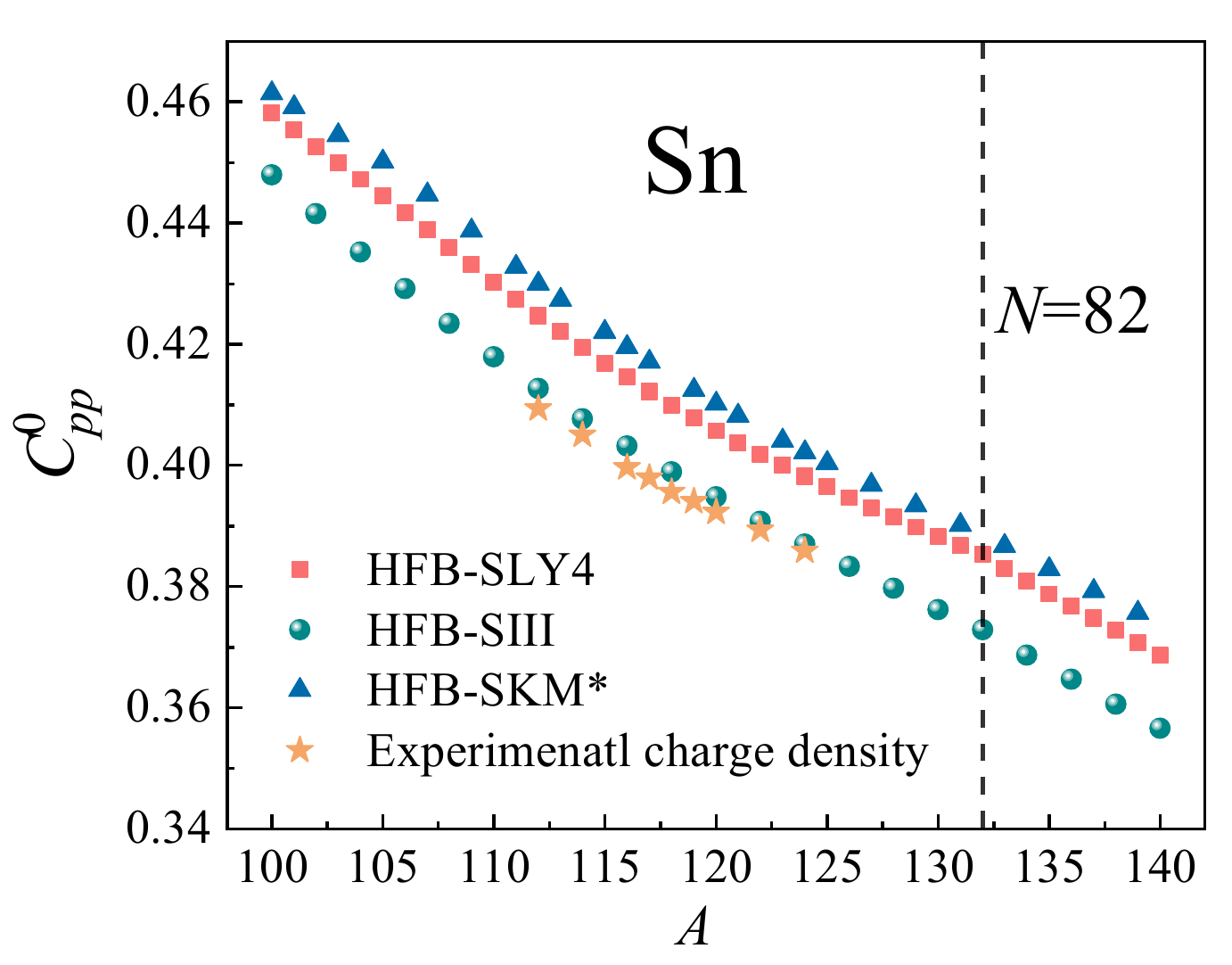}
    \caption{The $pp$ contacts $C^{0}_{pp}$ for Sn isotopes as a function of mass number $A$ varying from 100 to 140, using the experimental charge density~\cite{DEVRIES1987} and the HFB point-proton density with SLY4, SIII, and SKM* parameter sets. The vertical dashed line marks the magic number $N=82$.}
    \label{fig:Sn_pp}
\end{figure}


Shown in Fig.~\ref{fig:Sn_pp} is the $pp$ contact $C^{0}_{pp}$ for Sn isotopes with the mass number $A$ varying from 100 to 140. We calculate $C^{0}_{pp}$ using both the point-proton density distribution calculated by the Skyrme HFB model and the experimental charge density distribution given by the elastic electron scattering~\cite{DEVRIES1987}. Using the HFB proton density distribution, the value of $C^{0}_{pp}$ exhibits a consistent decrease as the mass number increases. Due to the existence of the tensor part of $NN$ force, the short-range $NN$ pairs are dominated by the $np$ pairs, which outnumber $pp$ pairs by approximately a factor of 20~\cite{Hen2014}. Consequently, with the increasing mass number, the correlated partner nucleon of a proton prefers a neutron, forming $np$ pairs. This results in the decreasing $pp$ contact shown in Fig.~\ref{fig:Sn_pp}. When the neutron number passes through the magic number 82, no obvious shell effects are observed. It is consistent with the former investigation that nuclear contact has no correlation with the shell effect.

\begin{figure}
    \centering
    \includegraphics[width=1\columnwidth]{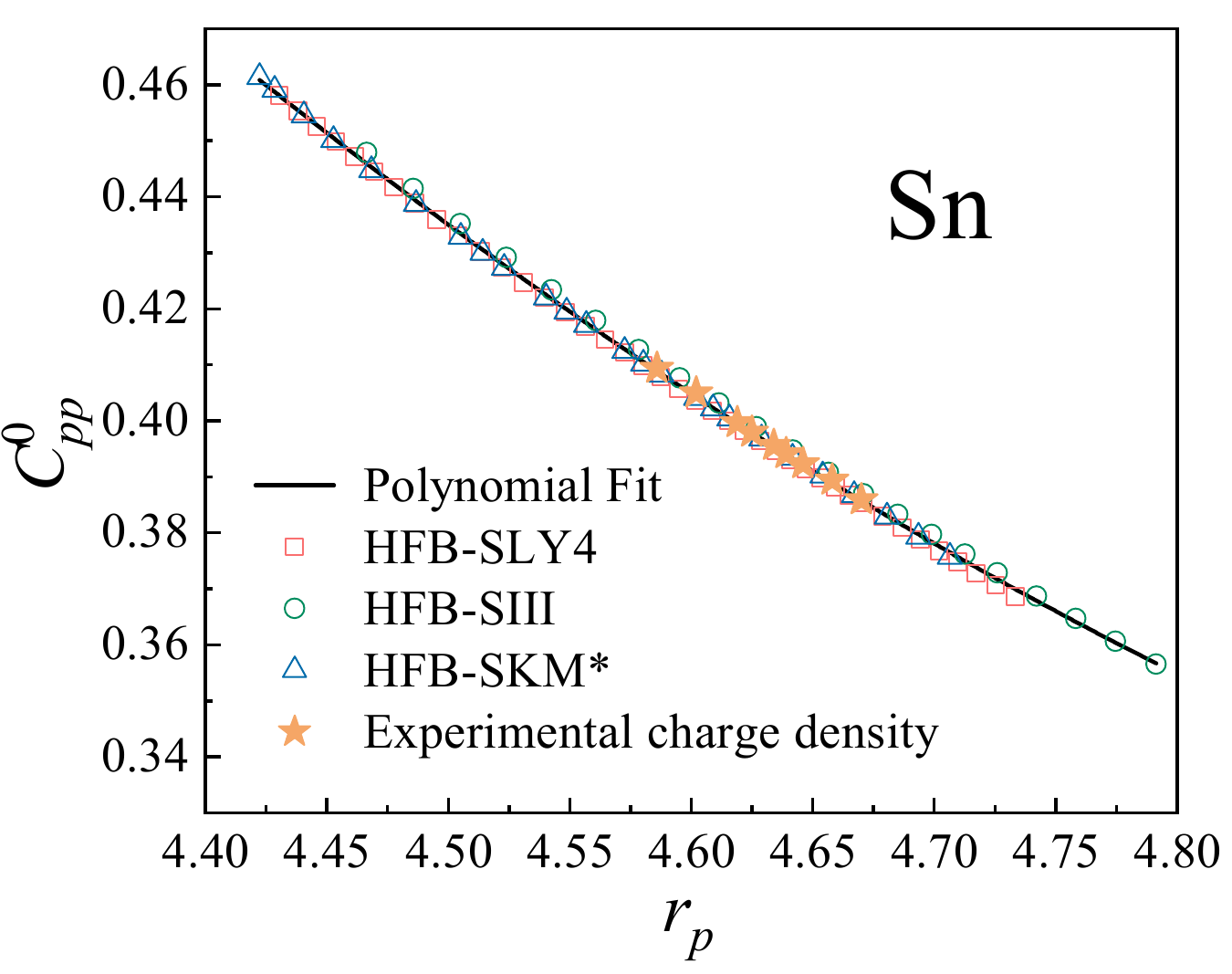}
    \caption{Correlation between the $pp$ contact $C^0_{pp}$ and the RMS radius of proton density distributions for Sn isotopes. The $pp$ contacts are calculated with the point-proton density distribution calculated by the Skyrme HFB model and charge density distribution from elastic electron scattering experiments~\cite{DEVRIES1987}. Note that the experimental charge density corresponds to the charge RMS radius. The polynomial fit is $C^0_{pp}=5.35611-1.86809 r_p +0.17212 r_p^2$ with a coefficient of determination $R^2=0.99902$.}
    \label{fig:correlation-Sn}
\end{figure}

In Ref.~\cite{Weiss2019}, authors fit a relation between $pp$ contact values and mass number within the form $C^0_{pp}(A)=bZ^2/A$, where $b$ is a fitting parameter and $Z$ is the proton number.
For different HFB proton densities or experimental charge densities, $pp$ contacts $C^{0}_{pp}$ show similar trends but different values, that is, the dependence of $C^{0}_{pp}$ on $A$ is some kind of model-dependent. 
Section~\ref{contact_theo} demonstrates that the $pp$ contact is derived from the point-proton density distribution and, therefore, has a close connection with the root-mean-square (RMS) radius of the proton density distributions $r_p$. Figure~\ref{fig:correlation-Sn} presents the relation between $C^0_{pp}$ and $r_p$, revealing a remarkable quadratic relation with a coefficient of determination $R^2=0.99902$. This high $R^2$ value indicates a strong, model-independent correlation between $C^0_{pp}$ and $r_p$. Additionally, the small coefficient of the quadratic term suggests a nearly linear relationship. Note that the experimental charge density corresponds to the charge RMS radius $r_C$, that is, the correlation is robust, regardless of whether $r_C$ or $r_p$ is used. This robustness makes it reasonable to extract $pp$ contacts with the experimental charge density, even though the derivation of $C^0_{pp}$ is based on the proton density.




The correlation shown in Fig.~\ref{fig:correlation-Sn} can respond well to the different $C^0_{pp}$ values in Fig.~\ref{fig:Sn_pp} using different HFB proton densities and the experimental charge density. Figure~\ref{fig:correlation-Sn} demonstrates that a larger radius leads to a smaller $C^0_{pp}$ value. 
It can be seen that for a chosen nucleus, larger $r_p$ (or $r_C$) values lead to smaller $C^0_{pp}$ values, satisfying the relationship as demonstrated in Fig.~\ref{fig:correlation-Sn}.





\begin{figure}
    \centering
    \includegraphics[width=1\columnwidth]{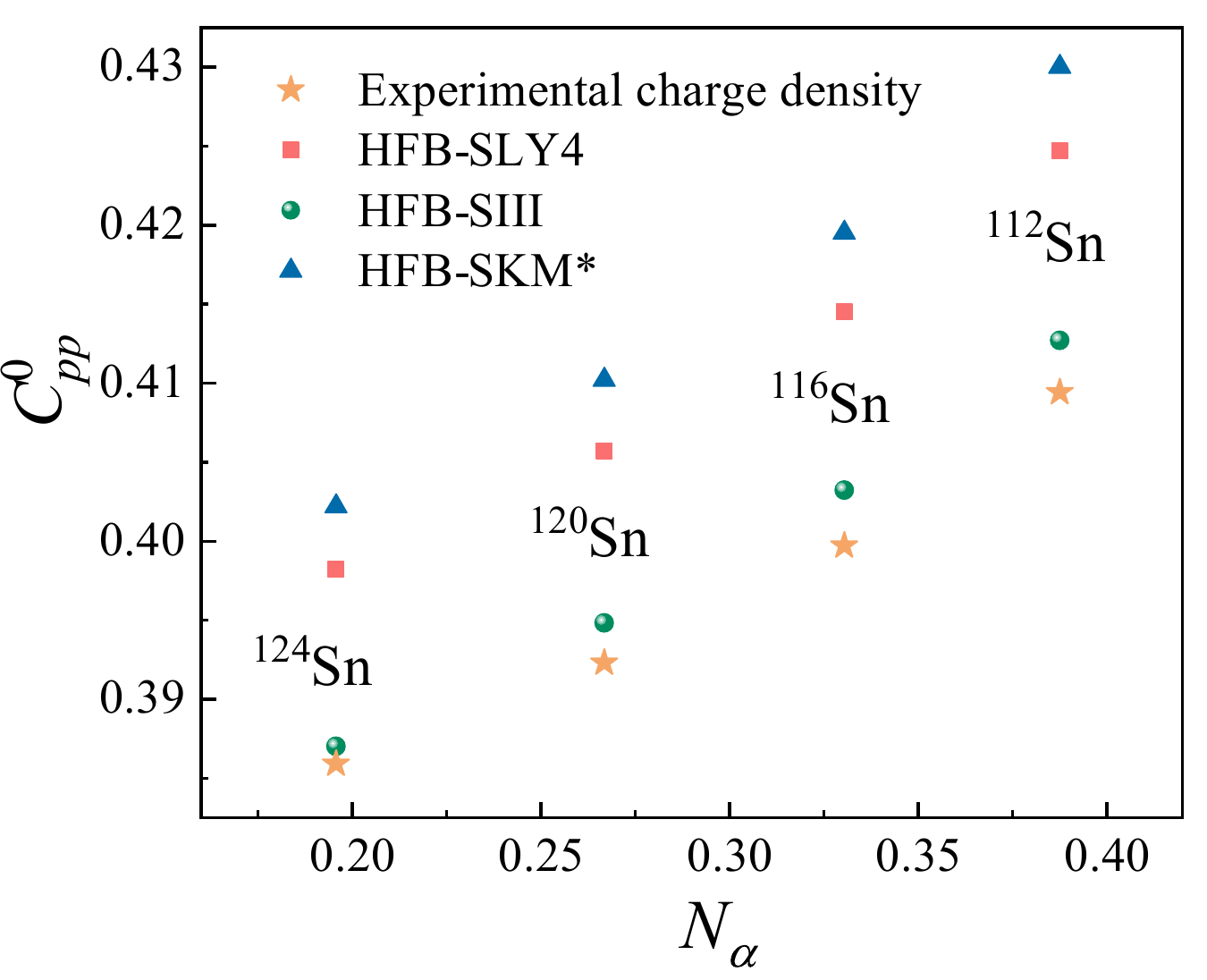}
    \caption{$pp$ contact $C^{0}_{pp}$ versus effective number of $\alpha$ clusters $N_{\alpha}$ for Sn isotopes. The $pp$ contacts are calculated with the point-proton density distribution calculated by the Skyrme HFB model and charge density distribution from elastic electron scattering experiments~\cite{DEVRIES1987}. The $N_{\alpha}$ values are extracted from the $(p,p\alpha)$ experiment.}
    \label{fig:Cpp-alpha}
\end{figure}

In a previous study, authors observed the reduction of $(p,p\alpha)$ reaction cross section with increasing mass number for Sn isotopic chains~\cite{Tanaka2021}. It reflects the decreasing effective number of $\alpha$ clusters $N_{\alpha}$, representing the probability of finding $\alpha$ clusters. This is consistent with the behaviour of short-range $pp$ pairs. To uncover the interplay between the $pp$ pairs and $\alpha$ clusters,  we plot the $pp$ contact $C^{0}_{pp}$ versus $N_{\alpha}$ in Fig.~\ref{fig:Cpp-alpha}. A consistent trend between $C^{0}_{pp}$ and $N_{\alpha}$ is observed for all $C^{0}_{pp}$ results using the HFB point-proton density and the experimental charge density. This implies a tight interplay between the formation of short-range $pp$ pairs and $\alpha$ clusters.


\begin{figure}
    \centering
    \includegraphics[width=1\columnwidth]{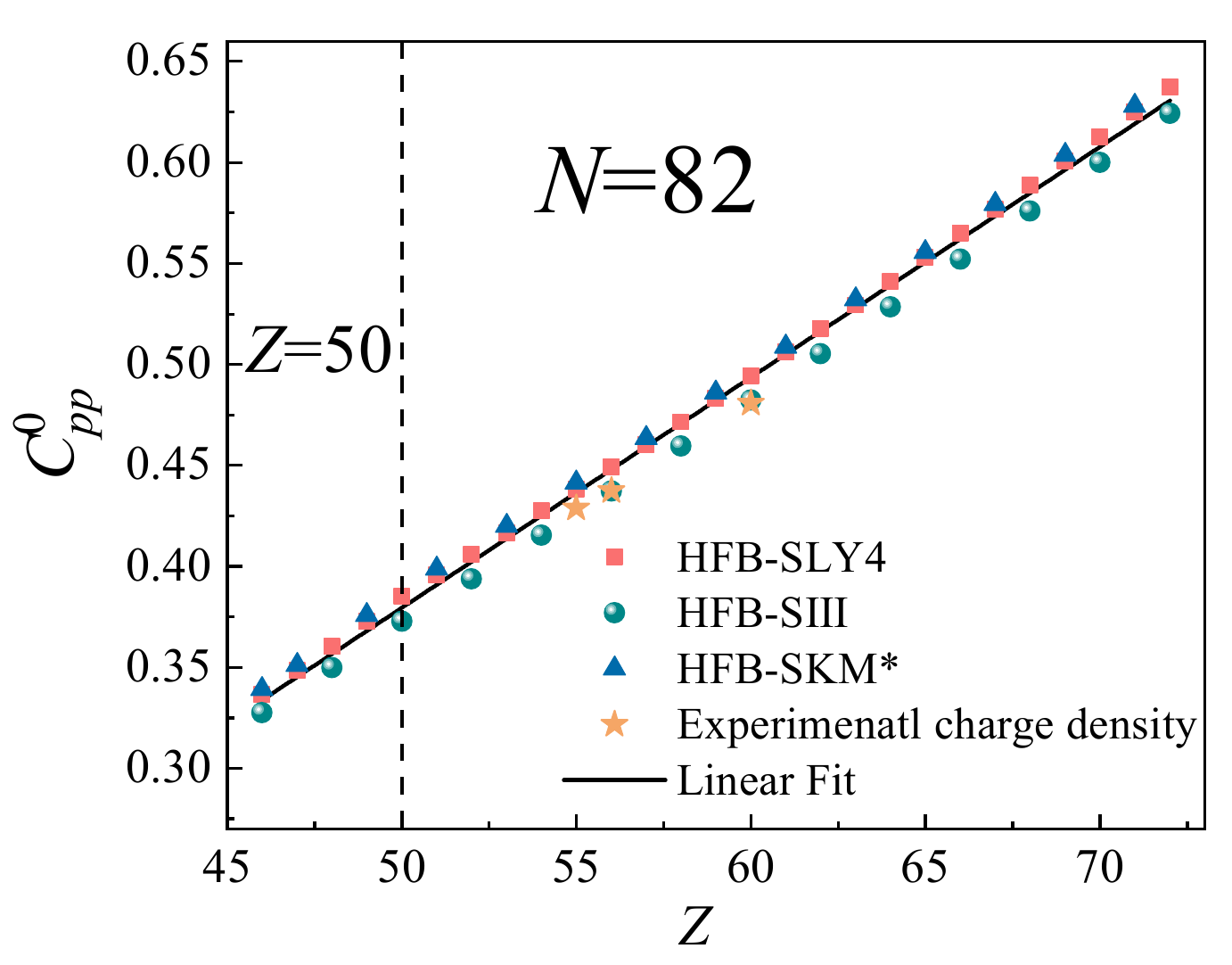}
    \caption{The $pp$ contacts $C^{0}_{pp}$ for $N=82$ isotones as a function of proton number $Z$ varying from 46 to 72, using the experimental charge density and the HFB point-proton density with SLY4, SIII, and SKM* parameter sets. The dashed line marks the magic number $Z=50$. The linear fit is $C^{0}_{pp}=-0.19089+0.01141A$ with a correlation coefficient $r=0.99742$.}
    \label{fig:82_pp}
\end{figure}

For $N=82$ isotones, the $pp$ contacts $C^0_{pp}$ are also calculated with both the point-proton density distribution calculated by the Skyrme HFB model and the experimental charge density distribution given by elastic electron scattering experiments. The $C^0_{pp}$ results versus proton number $Z$ are presented in Fig.~\ref{fig:82_pp}. The experimental charge density distributions are available for 
$^{136}$Xe, $^{138}$Ba, and $^{142}$Nd~\cite{DEVRIES1987}. 
A significant linear trend between $C^0_{pp}$ and $Z$ is found with a correlation coefficient $r=0.99742$. 
Several quasi-elastic electron scattering experiments have been carried out to determine the $np$ and $pp$ SRC pairs, suggesting a basically consistent fraction of $np$ and $pp$ SRC pairs in various nuclei~\cite{Subedi2008}. This conclusion is verified by the linear correlation between $C^0_{pp}$ and $Z$ shown in Fig.~\ref{fig:82_pp}. 
With the increasing $Z$ for $N=82$ isotones, protons can form new $pp$ pairs or new $np$ pairs. The fixed $np$-to-$pp$ fraction leads to the uniform increase of both $np$ and $pp$ SRC pairs, that is, the $C^0_{pp}$ value increases linearly versus $Z$. The linear correlation, $C^0_{pp}\propto Z$, was also observed in Ref.~\cite{Weiss2015} by analyzing the nuclear contact in the momentum space. 


Quasi-elastic electron scattering experiments, while being a direct and effective method for determining nuclear contact, are currently limited to stable nuclei and are not applicable to unstable nuclei.
The generalized contact formalism allows us to extract the $pp$ contact $C^0_{pp}$ with the charge density distribution. Moreover, benefiting from the development of radioactive nuclear beam facilities, elastic electron scattering off short-lived unstable nuclei is available at RIKEN. The experiment has been performed successfully with online-produced $^{137}$Cs~\cite{Tsukada2023}. They are now focusing on the elastic electron scattering around $^{132}$Sn region~\cite{Ohnishi2023}.  This makes available the experimental extraction of $C^0_{pp}$ for unstable nuclei.

\section{conclusions}
\label{Concl}
By combining the generalized contact formalism and the Skyrme Hartree-Fock-Bogolyubov model, we systematically investigated the nuclear $pp$ contact, which is proportional to the $pp$ pair number, for Sn isotopes and $N=82$ isotones. 
Our results reveal the dominance of $np$ pairs and a consistent number of $pp$ pairs across different nuclei. 
The consistent decreasing trend of $pp$ contact is similar to the effective number of $\alpha$ clusters along the Sn isotopic chain, suggesting a tight interplay between the formation of the $pp$ pairs and $\alpha$ clusters.
Moreover, we identify a model-independent nearly linear correlation between the $pp$ contact and the RMS radius of proton densities, which holds for both stable and unstable Sn isotopes. 

With the radioactive nuclear beam facilities at RIKEN, elastic electron scattering experiments, and thus the $pp$ contact, are feasible for unstable nuclei.
Our studies provide a more complete picture of nuclear contact and are expected to be helpful for exploring the correlation between short-range correlations and other fundamental nuclear properties. 

\begin{acknowledgments}
This work is supported by the National Natural Science Foundation of China (Grants No.\ 12035011, No.\ 11975167,  No.\ 11947211, No.\ 11905103,  No.\  11881240623, No.\ 11961141003, and No.\ 12375122), and by the National Key R\&D Program of China (Contracts No.\ 2023YFA1606503 and No.\ 2018YFA0404403). 
\end{acknowledgments}

\normalem
\bibliography{references}
\end{document}